# Investigating Safety Impacts of Roadway Network Features of Suburban Arterials in Shanghai, China

**Xuesong Wang, PhD (Corresponding Author)**
Professor
School of Transportation Engineering, Tongji University
4800 Cao'an Road, Jiading District
Shanghai, 201804, China
Phone: +86-21-69583946
Fax: +86-21-65982897
Email: wangxs@tongji.edu.cn

**Jinghui Yuan**
School of Transportation Engineering, Tongji University
4800 Cao'an Road, Jiading District
Shanghai, 201804, China
Phone: +86-21-69583946
Fax: +86-21-65982897
Email: yuanjinghui718@gmail.com

**Grant G. Schultz, PhD**
Professor
Department of Civil & Environmental Engineering, Brigham Young University
368 Clyde Building, Provo
UT 84602, USA
Phone: (801) 422-6332
Fax: (801) 422-0159
E-mail: gschultz@byu.edu

**Wenjing Meng**
Suzhou Singapore International School
208 Zhongnan Street, Suzhou, 215000, China
Phone: +86-0512-65330988
Email: mengwenjing.anna@gmail.com

**ABSTRACT**

With the rapid changes in land use development along suburban arterials in Shanghai, there is also a corresponding increase in traffic demand along these arterials. With a preference toward increased accessibility and efficiency, these arterials have been installed with an increased number of signalized intersections and accesses to serve local traffic needs. The absence of a defined functional hierarchy along the road network, together with the non-uniform installation of signals and accesses tends to deteriorate arterial safety. Previous studies on arterial safety have generally been based on a single type of road entity (either intersection or roadway segment). These studies only analyzed partial safety impacts of signal spacing and access density, as these factors would significantly influence the safety performance of both intersections and roadway segments. Macro level safety modeling was usually applied to investigate the relationship between the zonal crash frequencies and demographics, road network features and traffic characteristics. In this study, a new modeling strategy was proposed to analyze the safety impacts of roadway network features (i.e., road network patterns, signal spacing and access density) of arterials by applying a macro level safety modeling technique. Bayesian Conditional Autoregressive models were developed for arterials covering 173 Traffic Analysis Zones in the suburban area in Shanghai. The results identified that the road network pattern with collector roads parallel to the arterials was shown to be associated with fewer crashes than those without parallel collectors. Higher signal density and access density also tended to increase crash frequencies on arterials.

# INTRODUCTION

With the rapid growth of the suburban area in Shanghai, the density of land use along suburban arterials is increasing at a dramatic rate due to a preference toward increased accessibility and efficiency of arterials. The urbanization rate of the Shanghai suburban area has exceeded 70% in 2010 compared to 47% in 2000 (*1*). There appears to be an increasing traffic demand along these suburban arterials and the frequent access of local traffic has uncovered several issues about the road network structure, signal spacing and access density.

In the rapid developing suburban area in China, the road network construction tends to lag behind the land use development. Many local streets connect directly to arterials with no collector roads between them. Without a properly planned functional hierarchy road network to distribute local traffic, more and more signals and accesses have to be installed, which tends to decrease the efficiency and safety of the arterials. To compound the problem there are no current authoritative guideline for the setting of signal spacing along suburban arterials in Shanghai. The addition of more non-uniform signal spacing tends to reduce arterial travel speeds thereby resulting in an excessive number of stops, which may increase the crash occurrence. High access density is commonly observed on suburban arterials which leads to increasing traffic conflicts. All of these problems about road network structure, signal spacing and access features are contributing to the deterioration of the safety performance of suburban arterials.

Previous studies on arterial safety have generally been conducted by modeling a single type of road entity, either intersection or roadway segment. While signal spacing and access density have safety impacts on both intersections and roadway segments, previous researchers have only estimated the partial safety effects of those factors on the part of arterials. The countermeasures development or safety benefits evaluation for those factors should be based on a clear understanding of the full safety impacts on both segments and intersections. The safety impacts of road network patterns on suburban arterials have rarely been investigated in previous research. The road network structure plays an important role in trip route determination and could aid in the distribution of local traffic for arterials. A functional hierarchy road network could reduce the access of local traffic along arterials, which may improve the safety performance of arterials by reducing the traffic conflicts.

Macro level safety modeling is conducted to investigate the relationship between the zonal crash frequency and social economic, demographics, road network features and traffic characteristics. In this study, a new modeling strategy was proposed by applying the macro level safety modeling method to investigate the full impacts of access density, signal spacing and road network patterns on arterial safety based on the traffic analysis zone (TAZ).

# LITERATURE REVIEW
## Road Network Patterns
Road network structure is one macro level variable that plays an important role in trip route determination and in overall network safety. The road network adjacent to the suburban arterials is highly correlated with the traffic operation characteristics and access features on the arterials. A functional hierarchy road network could distribute the local traffic for the arterials, resulting in a reduction of traffic conflicts on arterials. The safety impacts of road network patterns on arterials have rarely been investigated in previous studies. Several studies have, however, analyzed the safety impacts of road network patterns at the zonal-level. Wang et al. (*2*) found that the sparse street pattern with sparse roadway layouts outperformed the other patterns (Mixed, Loops and Lollipop and Parallel and Grid) in zonal safety. Marshall and Garrick (*3*) found that the increased street connectivity in the form of the link to node ratio was significantly associated with more crashes.

## Signal Spacing
Signal spacing refers to the distance between two consecutive signalized intersections. Two main features are derived from the signal spacing: signal density and uniformity of signal spacing. The safety impacts of signal density were consistent in the literature, indicating that higher signal density leads to more crashes (*4; 5; 6; 7*). Abdel-Aty and Wang (*7*) analyzed the safety impacts of signal spacing on intersections. They found that longer signal spacing was associated with fewer crashes. Wang et al. (*4*) developed several hierarchical Bayesian models to investigate the safety impact of signal spacing based on the segment-level data collected from eight suburban arterials in Shanghai. They found that one more signalized intersection installed per kilometer was associated with an increase in crash frequency of 172%.

Non-uniform signal spacing will disrupt traffic operation and make drivers frequently accelerate and decelerate. The subsequent increase in speed variation decreases safety. Gluck et al. (*6*) reported that long and uniform signal spacing could achieve efficient traffic operations and improve safety. Wang et al. (*4*) investigated the safety impacts of signal uniformity at the arterial level. They found that the larger standard deviation of signal spacing tended to increase the minor injury crash frequencies.

## Access Density
Many studies have shown that access density is one of the key factors that influence crash frequencies. Gluck et al. (*6*) reported that doubling the number of access points on an arterial roadway from 6.25 to 12.5 per kilometer would increase crash rates by 30 to 40%. Dinu and Veeraragavan (*8*) classified the accesses into private access points (houses/shops) and public access points (driveways) to investigate the safety impacts of various access related factors on two-lane undivided rural highways in India. They found that the increase in the number of both private and public access points resulted in increased crash frequencies.

Several before and after analyses on the safety impacts of access density can also be found in the literature. For example, Schultz et al. (*9*) conducted before and after analysis on arterials in Utah. They found that the crash frequency is generally expected to increase with the density of accesses on arterial segments.

**Modeling Methodology**
The majority of previous studies on arterial safety were conducted for a single type of road entity, that is, either intersection or road segment. Abdel-Aty and Wang (*7*) investigated the significant safety influence factors for the intersections along corridors based on the intersection functional areas within 250 ft (76.25 m) of the center of the intersection. Wang et al. (*4*) conducted the safety analysis for arterials based on the segments divided by signalized intersections excluding functional area of the intersection. While signal spacing and access density influence the crash frequencies of both segments and intersections, the partial safety impacts on road segments or intersections could not reveal the real relationship between such factors and arterial safety.

Several macro level safety analyses were conducted at the TAZ level to investigate the relationship between the TAZ level crash frequencies and demographics, roadway network features and traffic characteristics (*2; 10; 11; 12*). A conventional Generalized Linear Model (GLM) was developed under the assumption that all samples are independent. Those assumptions are sometimes violated because the TAZs in close proximity to each other are similar in nature as well as in safety performance. To address this spatial correlation problem, numerous studies have been conducted over the years using a Conditional Autoregressive (CAR) model (*2*).

## DATA PREPARATION
**TAZ Delineation and Macro Features Extraction**
This study was conducted based on the suburban area of Jiading and Baoshan Districts in Shanghai. According to the guidelines for TAZ delineation (*13*), TAZs are delineated by combining the river, town boundary, regional boundary, highway route map and railway route map using ArcGIS®. In generating the initial TAZ shape it was noted that several TAZs appeared to be an irregular shape and the land use type was not consistent within one TAZ. To address these problems, the initial TAZs were modified according to these rules: employing the main roads to act as the TAZ boundary (arterials take precedence over collector roads); the TAZ area must exclude rivers; keep the same land use type in each TAZ; and keep a regular shape for each TAZ. Following the procedure a total of 202 TAZs were delineated in the area of Jiading and Baoshan Districts.

The number of trip productions and attractions per day within each TAZ was collected from the Macro Travel Demand Model of Shanghai maintained by the local government. The land use properties distribution of Jiading and Baoshan Districts were acquired from the website of the Municipal Land Administration Bureau. All the land use types were classified into 7 classes: industrial (18.3%), commercial (22.3%), educational (7.4%), technical (8.4%), residential (20.8%), greenspace (8.4%) and agricultural (14.4%).

**Extraction of Road Features**

In this study, road features mainly include the arterial length, signal density and access density along arterials. Suburban arterials for this study were comprised of arterial highways and secondary highways. They were selected by the attribute of road grade using ArcGIS®. All of the suburban arterials of Jiading and Baoshan Districts are shown as Figure 1. The arterial length was calculated in the total mileage of arterials in every TAZ using the spatial join toolbox in ArcGIS®. Similarly, the total mileage of all roadway types could be calculated, and the road density of each TAZ could then be calculated by dividing by the corresponding area. The roadways on the boundary of TAZs were evenly distributed to the adjacent TAZs.

The number of signalized intersections and access points along arterials within each TAZ was calculated and then divided by the arterial length to generate the corresponding signal density and access density. As for the intersections that lie on the boundary of TAZs, they were allocated to the adjacent TAZs evenly.

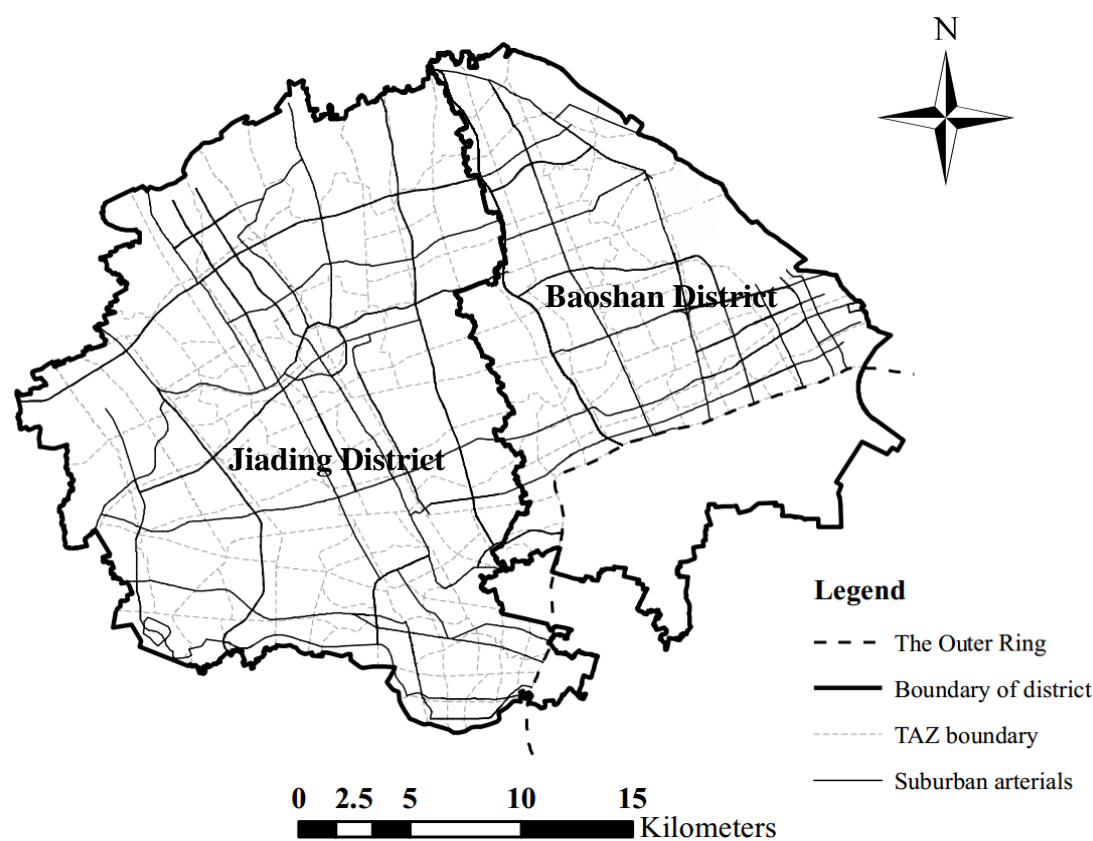

**FIGURE 1 Suburban arterials in Jiading and Baoshan districts (outside the outer ring).**

**The Betweenness Centrality and Road Network Graphical Patterns**
*Calculation of Betweenness Centrality*
The betweenness centrality is used to examine the shortest distance of any two nodes as a way to measure their importance in terms of traversing the network. It is based on the idea that a node is central if it lies between many other nodes, in the sense that it is traversed by many of the shortest paths connecting couples of nodes. The betweenness centrality of node *i* is defined as outlined in Equation 1(*14*):

$$C_i^B = \frac{1}{(N-1)(N-2)} \times \sum_{j \neq k \neq i} \frac{n_{jk}(i)}{n_{jk}} \tag{1}$$

Where $N$ is the total number of nodes, $n_{jk}$ is the number of shortest paths between node *j* and k, and $n_{jk}(i)$ is the number of shortest paths between *j* and k that contain node *i*.

For the zonal road network, the betweenness centrality is defined in Equation 2:

$$C^B = \frac{\sum_{i=1}^{N}(C_{i*}^B - C_i^B)}{N^3 - 4N^2 + 5N - 2} \tag{2}$$

Where $C_{i*}^B$ is the maximal value of $C_i^B$, the other parameters are the same as the above function.

The betweenness centrality was calculated as follows: extract the road network for each TAZ respectively and calculate the adjacent road network matrix for each TAZ using a customized functionality. The adjacent matrix for each TAZ was imported into UCINET, which is a social network analytical program that can calculate some network structural indices (*15*). The betweenness centrality of the road network within each TAZ could be calculated by the program.

*Relationship between Road Network Graphical Patterns and Betweenness Centrality*
In order to illustrate the relationship between the value of betweenness centrality and road network graphical patterns, a visual inspection was conducted on each TAZ considering the road network pattern and the value of betweenness centrality. Based on the inspection, a significant difference in the road network pattern for various values of betweenness centrality was found to exist. A comparison of road network pattern was conducted on 10 TAZs with the highest difference among their values of betweenness centrality. These TAZs were grouped by two categories to represent the lower and higher value of betweenness centrality separately as shown as Figure 2.

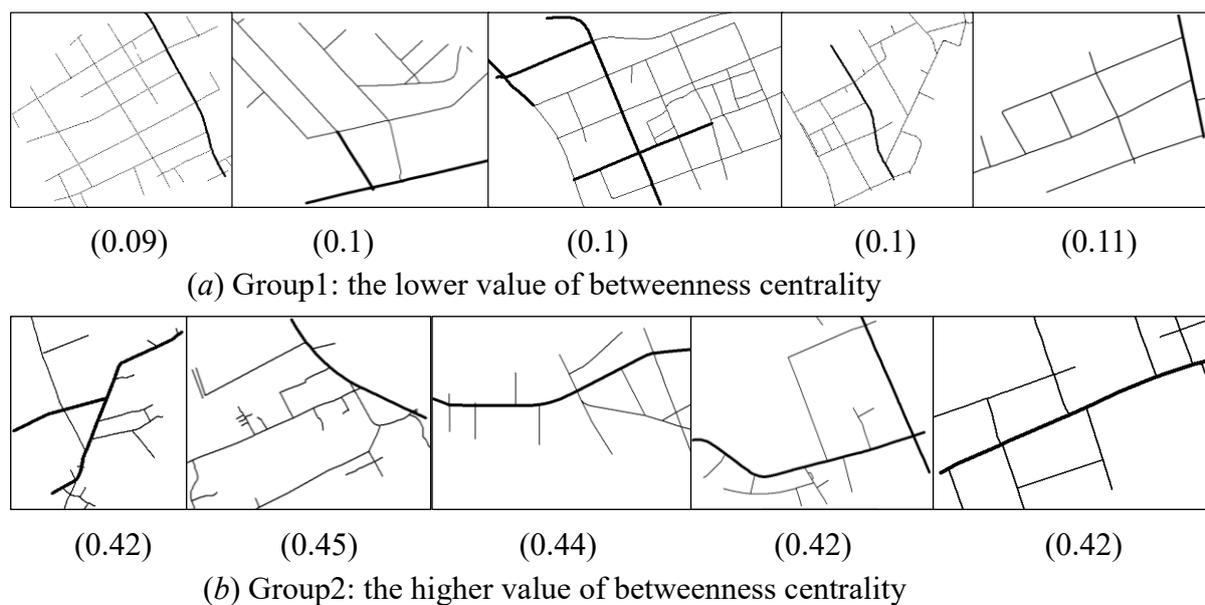

(a) Group1: the lower value of betweenness centrality

(b) Group2: the higher value of betweenness centrality

**FIGURE 2 Road network comparison of discrepant value of betweenness centrality (the bold lines represent suburban arterials).**

As can be seen from Figure 2, in Group1, there are one or more collector roads parallel and within a short distance of the arterials, and the entire road network shows a tendency of spreading perpendicular to the arterials. As a result, the local traffic within the TAZ would be distributed by the collector roads and local streets, and may not exhibit as much dependency on the arterials. In Group2, there seems to be more accesses along arterials compared to Group1, and there are rarely collector roads parallel with arterials or the collector roads are discontinuous. The road networks in Group2 are spread along arterials, showing tree-like structures with a higher dependency on the arterials, thus a higher betweenness centrality.

*Classification of Road Network Patterns*
Since betweenness centrality is an abstract continuous variable based on the topological features of the network structure, the effect of betweenness centrality on safety may be very difficult to understand and it may also be hard to identify which road network pattern can be referenced by a specific value of betweenness centrality. In order to investigate any intuitive safety impacts of road network patterns on arterial safety, the 173 TAZs were classified into four categories by the value of betweenness centrality combined with the road network graphical features within each TAZ. A summary of the four categories TAZ road network patterns is shown as Table 1.

**TABLE 1 Summary of Road Network Patterns and Betweenness Centrality**

| Road Network Patterns | Grid | Irregular Grid | Mixed | Lollipops |
|---|---|---|---|---|
| Roadway network figures | 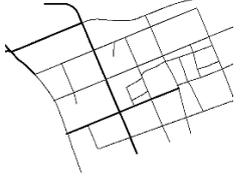 | 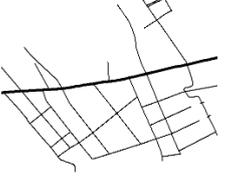 | 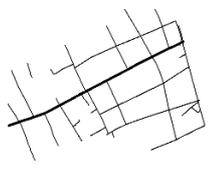 | 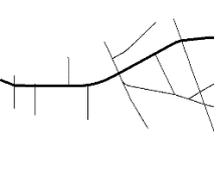 |
| Description | Grid structure with collector roads parallel to arterials | Grid structure with discontinuous collector roads parallel to arterials | Mixed structure of grid and tree with discontinuous collector roads parallel to arterials | Tree structure without continuous collector road parallel to arterials |
| Sample size | 29 (17%) | 92 (53%) | 34 (20%) | 18 (10%) |
| Betweenness Centrality | [0.01, 0.14] | [0.15, 0.29] | [0.3, 0.39] | [0.4, 0.65] |

**Crash Data**

After all arterials within the study area were identified, 29 TAZs were excluded from the dataset because they did not contain any suburban arterials. The modeling data were collected from the remaining 173 TAZs. Total crashes on the arterials of Jiading and Baoshan Districts in the year 2012 were collected for analysis. The geocoding procedure in ArcGIS® was used to locate the crashes on the GIS base map according to the location description in the crash reports. The TAZ shape was overlaid on the base map and then the crash numbers that occurred on the arterials of each TAZ were calculated using the spatial join toolbox in ArcGIS®.

The trip productions and attractions, arterial length, signal density, access density and road density were calculated for each TAZ. The descriptive statistics of these independent variables are listed in Table 2

**TABLE 2 Descriptive Statistics of Independent Variables at TAZ-level**

| Variables | Description | Mean | Min | Max | SD |
|---|---|---|---|---|---|
| TAZ area (km$^2$) | The area of each TAZ | 3.26 | 0.75 | 13.42 | 2.4 |
| Ln_Production | Natural log of trip productions per day within each TAZ | 9.89 | 7.13 | 12.1 | 0.88 |
| Ln_Attraction | Natural log of trip attractions per day within each TAZ | 9.84 | 6.82 | 12.02 | 0.96 |
| Arterial length (km) | The total length of arterials within each TAZ | 3.13 | 0.34 | 7.51 | 2.05 |
| Access density | Number of accesses per kilometer along arterials | 2.08 | 0.47 | 7.73 | 1.31 |
| Signal spacing | Number of Signalized intersections per kilometer along arterials | 1.74 | 0.54 | 4.05 | 0.75 |
| Road density (km/ km$^2$) | Length of road per square kilometer of TAZ area | 3.11 | 0.29 | 12.63 | 2.13 |

## MODELING STRATEGIES

In this study, a new strategy was proposed to analyze the safety impacts of suburban arterials at the TAZ level. The crash frequencies of the arterials within each TAZ were analyzed as the dependent variable. The roadway network features (i.e., road network patterns, signal spacing and access density) and traffic characteristics were analyzed as the independent variables. Through this strategy, the full impacts of signal density and access density on arterial safety (including road segments and intersections) could be investigated. In addition, the safety impacts of road network patterns within each TAZ could be taken into consideration, while the traditional strategies do not. An illustration of the comparison between different modeling strategies is shown in Table 3.

**TABLE 3 Modeling Strategy Illustration**

| Modeling Strategy | Models | Figure Illustration | Dependent Variables | Independent Variables |
|---|---|---|---|---|
| Traditional approach | Road segment model | 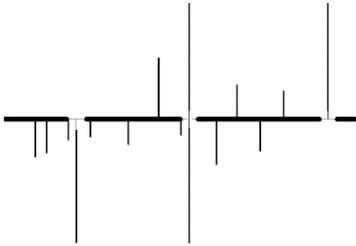 | Crashes occurred on the road segments | Traffic characteristics, Roadway features |
| | Intersection model | 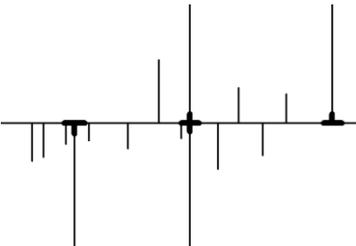 | Crashes occurred in the functional area of intersections | Traffic characteristics, Roadway features |
| New strategy | Model for arterials at TAZ-level | 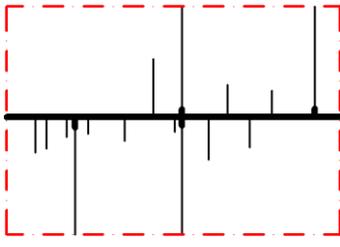 | Total crashes occurred on the arterials within the TAZ | Trip generation, Road network structures, Roadway features, |

## STATISTICAL MODELING METHOD

### Bayesian Poisson Lognormal CAR Model

The Bayesian Poisson Lognormal CAR model was proposed for this analysis. The Bayesian inference has the advantage of mitigating the estimating error due to the randomness of each crash occurrence. All of the parameters are regarded as random variables that are characterized by a prior distribution and they are estimated after combining the prior distribution and the sample data. The theoretical framework for Bayesian inference can be expressed as outlined in Equation 3:

$$\pi(\theta|y) = \frac{L(y|\theta)\pi(\theta)}{\int L(y|\theta)\pi(\theta)d\theta} \quad (3)$$

where $y$ is the vector of observed data, $\theta$ is the vector of parameters required for the likelihood function, $L(y|\theta)$ is the likelihood function, $\pi(\theta)$ is the prior distribution of $\theta$, $\int L(y|\theta)\pi(\theta)d\theta$ is the marginal distribution of observed data, and $\pi(\theta|y)$ is the posterior distribution of $\theta$ given $y$.

The Poisson-lognormal CAR model was proposed in the Bayesian framework for this study to overcome this spatial correlation between the crash occurrences on arterials in neighboring zones. Letting $y_i$ represent crash frequencies occurring on the arterials within $TAZ_i$, it is assumed that the dependent variable $y_i$ follows the Poisson distribution as outlined in Equation 4:

$$y_i \sim Poisson(\lambda_i) \qquad (4)$$

where $\lambda_i$ is the expectation of $y_i$.

A random effect term $\theta_i$ and a spatial correlation term $\phi_i$ were introduced into the basic Poisson model to explain the site-specific heterogeneity and the spatial correlation between the neighboring TAZs separately. The model is defined as outlined in Equation 5 and Equation 6.

$$log(\lambda_i) = \psi_i = X'\beta + \theta_i + \phi_i \qquad (5)$$

$$\theta_i \sim Normal(0, \sigma_\theta^2) \qquad (6)$$

where $X'$ is the covariate matrix, $\beta$ is the vector of regression coefficients, $\theta_i$ is the random effect term, $\sigma_\theta^2$ is the variance of $\theta_i$, $\phi_i$ denotes the spatial correlation for $TAZ_i$, which is specified with a CAR prior. The conditional distribution of CAR prior is defined as outlined in Equation 7.

$$\phi_i | \phi_{(-i)} \sim N(\sum_j \frac{w_{i,j}}{w_{i+}} \phi_j, \frac{1}{\tau_c w_{i+}}) \qquad (7)$$

where $\phi_{-i}$ is collection of all $\phi$ except for $\phi_i$, $\tau_c$ is the precision parameter which accounts for the variation of the spatial dependence, $w_{i,j}$ is the entry on the proximity matrix and generally reflects the spatial relationship between TAZs $i$ and $j$, $w_{i+}$ is the sum of $w_{i,j}$ in the TAZs that is adjacent to $TAZ_i$. The proportion of variability in the random effects that is due to spatial autocorrelation (α) is defined as outlined in Equation 8.

$$\alpha = \frac{sd(\phi)}{sd(\theta) + sd(\phi)} \qquad (8)$$

where $sd$ is the empirical marginal standard deviation function.

**Spatial Proximity Structures**

An important detail associated with the CAR model is the weight matrix of the neighboring structures. Generally, researchers would use the 0-1 first order neighboring structure between spatial units to act as the weight matrix of neighboring zones hypothetically putting equal weights on the adjacent zones (*16; 17*). In this study, three types of spatial proximity structures were constructed, including 0-1 first order adjacency, common boundary length and total lane number of connecting arterials. The spatial correlation among TAZs can be expressed by a proximity matrix $W$ with entry $w_{i,j}$ indicating the spatial relationship between TAZs i and j. For example, the 0-1 first order adjacency was used to complete this weight matrix $W_1$, defined as outlined in Equation 9:

$$w_{i,j} = \begin{cases} 1, & \text{if } TAZ_i \text{ and } TAZ_j \text{ are adjacent} \\ 0, & \text{if } TAZ_i \text{ and } TAZ_j \text{ are not adjacent} \end{cases} \quad (9)$$

As to the other two kinds of spatial proximity, $w_{i,j}$ is equal to the common boundary length and total lane number of those arterials connecting the adjacent $TAZ_i$ and $TAZ_j$ respectively.

**Model Comparisons**

The Deviance Information Criterion (DIC) can be used to compare complex models by offering a Bayesian measure of model fitting and complexity (*18*). DIC is defined as outlined in Equation 10:

$$DIC = \overline{D(\theta)} + p_D \quad (10)$$

where $D(\theta)$ is the Bayesian deviance of the estimated parameter, and $\overline{D(\theta)}$ is the posterior mean of $D(\theta)$. $\overline{D(\theta)}$ can be viewed as a measure of model fitting, while $p_D$ denotes the effective number of parameters and indicates the complexity of models.

Models with smaller DIC are preferred. Very roughly, difference of more than 10 might definitely rule out the model with the higher DIC. Differences between 5 and 10 are considered substantial (*19*).

In addition, the R-square is used to estimate the goodness of the predictive performance, it is calculated according to Equation 11.

$$R^2 = 1 - \frac{\sum_{i=1}^{n}(y_i - \hat{y}_i)^2}{\sum_{i=1}^{n}(y_i - \bar{y}_i)^2} \quad (11)$$

Where $\hat{y}_i$ is the predictive crash frequency of $TAZ_i$, $n$ is the total number of TAZs. $\bar{y}_i$ is the global mean of crash frequencies.

# MODELING RESULTS

The Bayesian method is usually implemented by using a Markov Chain Monte Carlo (MCMC) algorithm. In this study, three Bayesian CAR models were developed using WinBUGS (*19*) to estimate the model parameters. The CAR prior was specified by the function of car.normal to reflect the spatial proximity relationship of the TAZs. Since there is no available historical data or results, the maximum likelihood estimation was performed by using SAS® with a negative binomial model to generate the informative priors. Means and variances of the independent variables were calculated and transformed into normally distributed prior distributions. The variance of random effect term follow inverse-gamma distribution ($10^{-3}$, $10^{-3}$) and the CAR precision parameter follow gamma distribution (0.1, 0.1).

For each model, two chains of 20,000 iterations were set. After the convergence (evaluated using the built-in Brooks-Gelman-Rubin (BGR) diagnostic statistic (20)), another 50,000 iterations were set to estimate the posterior distribution of the parameters. The estimation results of three models are summarized in Table 4

**TABLE 4 Model Comparison and Parameter Estimation Results**

| Variables | Model 1 Mean (95% BCI) | Model 2 Mean (95% BCI) | Model 3 Mean (95% BCI) |
|---|---|---|---|
| Intercept | **2.361 (2.236, 2.482)** | **2.353 (2.229, 2.475)** | **2.349 (2.219, 2.477)** |
| Ln_Production | **0.073 (0.044, 0.104)** | **0.075 (0.046, 0.1)** | **0.075 (0.048, 0.1)** |
| Ln_Attraction | **-0.086 (-0.116, -0.058)** | **-0.085 (-0.11, -0.059)** | **-0.085 (-0.11,-0.059)** |
| Arterial length | **0.177 (0.167, 0.186)** | **0.178 (0.169, 0.187)** | **0.176 (0.167, 0.185)** |
| Access density | **0.107 (0.096, 0.118)** | **0.107 (0.096, 0.118)** | **0.107 (0.096, 0.118)** |
| Signal spacing | **0.314 (0.29, 0.338)** | **0.315(0.291, 0.338)** | **0.315 (0.291, 0.339)** |
| Road density | **-0.027 (-0.036, -0.019)** | **-0.027 (-0.035, -0.019)** | **-0.028 (-0.036, -0.02)** |
| **Road network patterns (Base: Grid)** | | | |
| Irregular Grid | **0.443 (0.382, 0.503)** | **0.433 (0.377, 0.491)** | **0.443 (0.386, 0.501)** |
| Mixed | **0.537 (0.47, 0.606)** | **0.546 (0.48, 0.611)** | **0.547 (0.482, 0.612)** |
| Lollipops | **0.692 (0.618, 0.767)** | **0.692 (0.621, 0.764)** | **0.689 (0.617, 0.761)** |
| **Land use types (Base: Industrial)** | | | |
| Commercial | **0.15 (0.105, 0.196)** | **0.155 (0.11, 0.199)** | **0.144 (0.01, 0.189)** |
| Educational | 0.024 (-0.048, 0.096) | 0.018 (-0.053, 0.088) | 0.025 (-0.046, 0.095) |
| Technical | **-0.115 (-0.184, -0.047)** | **-0.117 (-0.184, -0.05)** | **-0.118 (-0.185, -0.05)** |
| Residential | **0.198 (0.152, 0.244)** | **0.195 (0.151, 0.241)** | **0.198 (0.153, 0.242)** |
| Greenspace | **-0.082 (-0.151, -0.011)** | **-0.078 (-0.146, -0.009)** | **-0.076 (-0.145, -0.007)** |
| Agricultural | 0.019 (-0.033, 0.071) | 0.018 (-0.033, 0.069) | 0.025 (-0.026, 0.077) |
| CAR effect ($\tau_c$) | **2.525 (1.95, 3.189)** | **0.034 (0.006, 0.131)** | **1.003 (0.627, 1.531)** |
| Random effect ($\frac{1}{\sigma_\theta^2}$) | **632.2 (70.82, 2413)** | **17.5 (11.3, 29.69)** | **82.28 (20.92, 449.4)** |
| $\alpha$ | **0.854 (0.721, 0.939)** | **0.366 (0.191, 0.543)** | **0.651 (0.527, 0.861)** |
| DIC | 1416.83 | 1420.51 | 1421.24 |
| $R^2$ | 0.774 | 0.778 | 0.776 |

*Note: The Mean (95% BCI) values marked in bold were significant at the 0.05 level.*
*Model 1: 0-1 first order adjacency; Model 2: Common boundary length; Model 3: Total lane number of connecting arterials.*

The significant CAR effect of the three models confirms the existence of spatial correlation for crash frequency among the neighboring TAZs. The value of $\alpha$ shows that the spatial correlation of Model 1 accounts for 85.4% of the overdispersion of the crash data, which is larger than that of the other two models. This indicates that the 0-1 first order neighboring structure is associated with better spatial correlation interpretation for the present dataset of analysis. The DICs for Model 1-3 are nearly equivalent, as the difference between them is lower than 5 (*19*). However, according to the predictive performance measured by R-square, Model 2 seems to have a slightly better predictive performance than the other two models.

The model comparison result indicates that Model 1 performs the best spatial explanation by employing the simplest model structure with the 0-1 first order adjacency. Generally, Model 1 should be selected as the best model based on the DIC results, even though it exhibits slightly worse predictive performance according to R-square.

## ANALYSIS OF RESULTS
### Road Network Patterns

All road network patterns variables were found to be significantly associated with crash occurrence on suburban arterials within each TAZ. According to the estimation coefficient, the TAZ with an irregular grid pattern road network tended to generate a 56% ($e^{0.443} - 1$) increase in crash frequency when compared with the grid pattern. The TAZ with mixed pattern road network tended to bring about 71% more crashes occurred on arterials when compared with the grid pattern. The TAZ with the lollipops pattern road network was expected to have twice number of crashes than the one with the grid pattern.

As the roadway network figures of each pattern in Table 1 show, the grid pattern road network contains several collector roads parallel to the arterials. All of the collector roads could distribute the local traffic and reduce the distribution of access traffic on the arterials, which may improve arterial safety. The irregular grid pattern is similar to the grid pattern but the collector roads are generally discontinuous, which tended to increase the crash frequencies when compared with the grid pattern with continuous collector roads. The mixed pattern is a mixed structure of tree and grid with discontinuous collector roads. The tree structure road network shows higher dependency on the arterials, and all the traffic should be severed by the arterials, which may result in an increasing traffic conflict on arterials. The lollipops pattern is nearly a whole tree structure that rarely includes collector roads parallel to arterials. It must perform with a higher dependency on arterials. This may result in a mixed traffic with high mobility vehicles and non-motorized vehicles sharing the same right of way, together with the frequent access traffic along arterials, which may significantly deteriorate the safety performance of arterials.

The box-plot shown in Figure 3 confirms the relationship between arterial crash frequencies and road network patterns. The TAZ with the road network more closely resemble tree structure and the absence of continuous collector road tends to increase the crash occurrence on arterials.

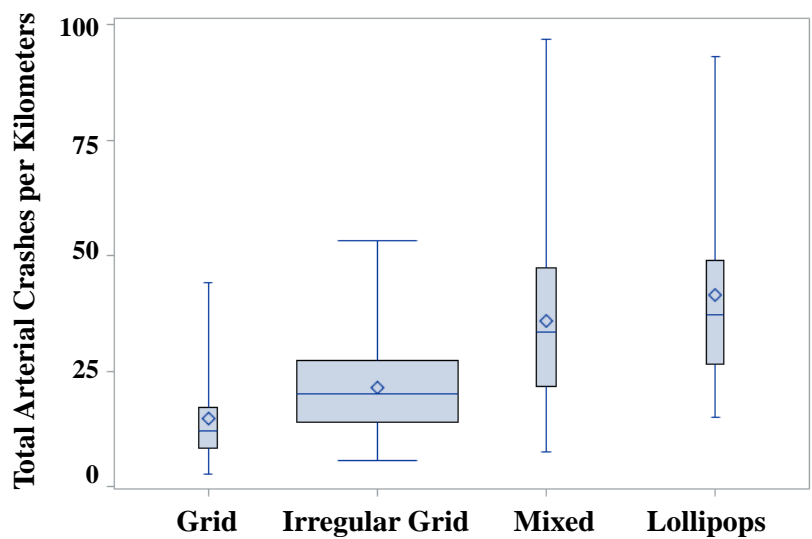

**FIGURE 3 Relationship between road network patterns and crash rate for the analyzed TAZs.**

**Signal Spacing**

In terms of signal spacing, the variable of signal density was found to have a positive effect on crash occurrence. Specifically, for every one kilometer of an arterial, one additional signal installation was associated with an increase in crash frequency of 36.9%. This can be explained in that the traffic flow may be more confusing due to the higher density of signalized intersections, and more lane-changing and overtaking behaviors appears to increase the risk of collision, which may lead to an increase in crash frequency. This is consistent with previous studies (*4; 5; 6; 8*). Wang et al. (*4*) found that the number of crashes per kilometer would predict an increase of about 172% with one more signal installed every one kilometer. The huge difference may be explained by the sample size, only 8 samples were collected for the signal density, which may result in a bias estimation.

**Access Density**

The coefficient of access density was significantly positive. The coefficient of 0.107 in Table 4 indicates that an increase of one access per kilometer would result in an increase of 11.3% ($e^{0.107} - 1$) in crash frequencies. This can be explained in that the random access of vehicles through these accesses along the arterials would result in more conflict points and deteriorate arterial safety. Several studies have analyzed the safety effect of access density and similar conclusions have been reached (*4; 5; 21*). Wang et al. (*4*) investigated the safety impacts based on the segment level and found that one more access installed per kilometer was associated with an increase of 10.3% in crash frequencies. This is a little bit smaller than our research, which can be explained in that the safety impacts on road segments and intersections were included simultaneously in our study. As shown in Figure 4, the significant growing tendency confirms the estimated result.

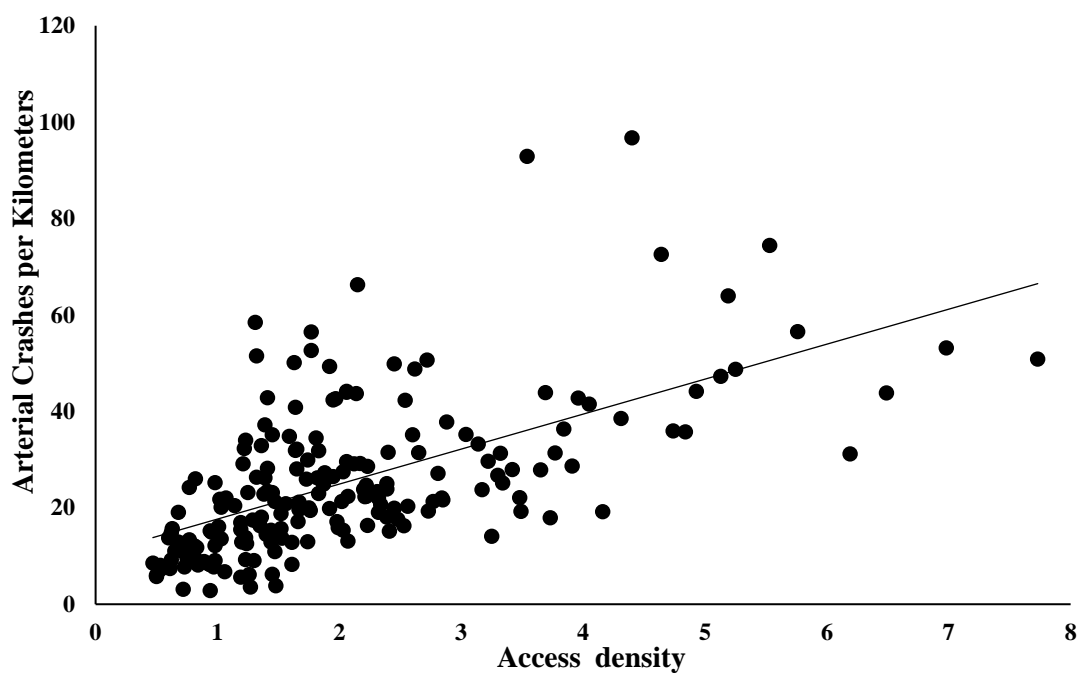

**FIGURE 4 Relationship of total arterial crash per kilometer and access density for each TAZ.**

The remaining variables of trip production and attraction, arterial length, road density and land use types were found to be significantly correlated with crash occurrence on arterials. More crashes tended to occur on the arterials within the TAZ with more trip production, less trip attraction, longer arterials and denser roadway network. The result also showed that more crashes occurred on commercial, technical, residential and greenspace land when compared to the industrial land.

## CONCLUSION AND DISCUSSION

This study attempted to identify the safety impacts of several safety factors (i.e. road network patterns, signal spacing and access density) on crash occurrence on arterials at TAZ-level by applying a macro level safety modeling method. The road network patterns were introduced into the safety analysis for suburban arterials. Bayesian Poisson-lognormal CAR models were developed with the modeling data collected from 173 TAZs delineated from the Jiading and Baoshan Districts (outside the Outer Ring) of Shanghai. Total crashes that occurred on the arterials within each TAZ were selected as the dependent variable. The model results showed that the CAR effects were significant in the three models, which confirmed the existence of cross-zonal spatial correlation in crash occurrence. The model comparison showed that the relatively best model is the one considering the proximity of neighboring zones by 0-1 first order adjacency.

According to the estimation results, several variables (i.e., trip production, trip attraction, arterial length, access density, signal spacing, roadway density, roadway network patterns and land use types) were shown to have statistically significant impacts on crash occurrence. Higher access density and signal density tended to increase the crash occurrence on arterials. The grid pattern road network with collector roads parallel to arterials outperformed the other road network patterns in arterial safety, and the tree-like lollipops pattern road network without continuous collector road parallel to arterials was shown to be the most dangerous. The denser roadway network was found to be correlated with fewer crashes. The commercial, technical, residential and greenspace land were associated with more crashes when compared to the industrial land.

The modeling results showed that the lollipops road network pattern was associated with twice number of crashes that occurred on arterials when compared to the grid pattern. Since the rapid development of suburban areas in China, the road network construction tends to lag behind, 10% of the road network pattern in Jiading and Baoshan Districts is the lollipops pattern. Due to the lack of a properly designed functional classification road network to distribute local traffic, more accesses have to be opened, which continues to decrease the efficiency and safety of arterials. It is recommended that the government plan the suburban area road network to more closely resemble the grid structures, not the tree structure with the arterials acting as the trunk. Several collector roads should also be constructed parallel to the arterials to improve the safety performance of suburban arterials.

Based on the results presented, the models developed for the suburban arterials appear to be useful with many applications such as setting standards for signal spacing and access density for suburban arterials and road network planning for the suburban area. With the improvement of data quality, the relationship between road network pattern and suburban arterial safety needs continued investigation.

## ACKNOWLEDGEMENT

This study was jointly sponsored by the Chinese National Science Foundation (No. 51522810) and the Science and Technology Commission of Shanghai Municipality (15DZ1204800).